\documentclass{article}
\usepackage[margin=0.5in]{geometry}
\makeatletter
\def\thanks#1{\protected@xdef\@thanks{\@thanks \protect\footnotetext{#1}}}
\makeatother
\date{}

\usepackage{graphicx}
\usepackage{mathabx}
\usepackage{pxfonts}
\usepackage{hyperref}

\newcommand{\R}{\mathbb{R}}
\newcommand{\mua}{\mu_{{\mathrm a}}}  
\newcommand{\mus}{\mu_{{\rm s}}'}
\newcommand{\rmi}{\mathrm{i}}

\begin{document}

\title{Utilising nanosecond sources in diffuse optical tomography}


\author{Meghdoot Mozumder$^{*,1}$, Jarkko Leskinen$^{1}$, Tanja Tarvainen$^{1}$\\ 
{\small $^{1}$Department of Applied Physics, University of Eastern Finland, P.O. Box 1627, 70211 Kuopio, Finland}\\
{\small $^{*}$Email: \href{mailto:meghdoot.mozumder@uef.fi}{meghdoot.mozumder@uef.fi}}}

\thanks{
This is the version of the article before peer review or editing, as submitted by an author to Measurement Science and Technology. IOP Publishing Ltd is not responsible for any errors or omissions in this version of the manuscript or any version derived from it. The Version of Record is available online at \href{https://iopscience.iop.org/article/10.1088/1361-6501/ac9e11/}{https://iopscience.iop.org/article/10.1088/1361-6501/ac9e11/}.}

\maketitle


\vspace{-8mm}
\begin{abstract}
Diffuse optical tomography (DOT) use near-infrared light for imaging optical properties of biological tissues. Time-domain DOT systems use pulsed lasers and measure time-varying temporal point spread function (TPSF), carrying information from both superﬁcial and deep layers of imaged target.

In this work, feasibility of nanosecond scale light pulses as sources for time-domain DOT is studied. Nanosecond sources enable using relatively robust measurement setups with standard analog-to-digital converter waveform digitizers, such as digital oscilloscopes. However, this type of systems have some properties, such as variations in source pulses and limited temporal sampling, that could limit their usage. In this work, these different aspects and possible limitations were studied with simulations and experiments.

Simulations showed that information carried by time-domain data of diffuse medium is on low frequencies. This enables usage of relatively slow response time measurement electronics, and image processing using Fourier-transformed time-domain data. Furthermore, the temporal sampling in measurements needs to be high enough to capture the TPSF, but this rate can be achieved with standard digital oscilloscopes. It was shown that, although variations in light pulses of nanosecond lasers are larger than those of picosecond sources, these variations do not affect signiﬁcantly on image quality. Overall, the simulations demonstrated the capability of nanosecond sources to be utilised in time-domain DOT in diffuse medium.

In this work, a prototype time-domain DOT experimental system utilising a high-energy nanosecond laser was constructed. The system is relatively robust consisting of a nanosecond Nd:YAG laser combined with optical parametric oscillator for light input and optical ﬁbres for guiding the light, and avalanche photodetector and high-bandwidth oscilloscope for TPSF measurements. The system was used in both absolute and difference imaging of two phantoms. The experiments veriﬁed that both absorbing and scattering objects can be reconstructed with good quality with time-domain DOT using a nanosecond laser.
\end{abstract}

%
\vspace{2mm}

\section{Introduction}\label{sect:intro}  

Diffuse optical tomography (DOT) is an imaging modality that uses visible red and/or near-infrared light for imaging spatially varying optical parameters in biological tissues \cite{arridge1999optical,gibson2005recent,durduran2015}. 
Distribution of these optical parameters provide tissue biochemical and structural information with applications, for example, in imaging of breast cancer, monitoring neonatal brain, functional brain imaging and pre-clinical small animal studies \cite{enfield2009optical,darne2014,durduran2015,grosenick2016,hoshi2016}. 

Experimental DOT systems can be divided into three types depending on the light source that is used.
These are continuous wave (CW), time-domain (TD), and frequency domain (FD) i.e. intensity modulated systems.
The CW sources enable fast data acquisition and usage of simple detectors \cite{schmitz2002instrumentation}. 
However, CW-DOT cannot distinguish between absorption and scattering effects unless a reference measurement to enable difference imaging is available \cite{arridge1998nonuniqueness,arridge1999optical}.
The FD-DOT systems use radio-frequency modulated light sources for illumination, and measure amplitude attenuation and phase delay of the transmitted light \cite{gibson2005recent}.
Time-domain diffuse optical tomography (TD-DOT) uses pulsed lasers for illuminating the tissues, and the time-varying boundary exitance, i.e. temporal point spread function (TPSF), is measured. \cite{pifferi2016new}.
Both FD and TD systems can be used for simultaneous absolute imaging of absorption and scattering distributions.
In DOT, the TD system are especially in the interest due to their capability to image through large thicknesses of tissue.
Furthermore, in TD-DOT, the information content of the measured TPSF is large, carrying information from both superficial and deep layers of the tissue \cite{gibson2005recent,pifferi2016new}.

Generally, the TD-DOT and near-infrared spectroscopy (NIRS) systems have been based on picosecond light sources.
{Wavelength-tunable light having picosecond or less duration can be generated using Ti:Sapphire and supercontinuum lasers \cite{selb2006time,contini2015,cooper2014monstir,lapointe2012}. 
These lasers operate in tens of MHz pulse repetition rates and can provide several watts (Ti:Sapphire) to several milliwatts (supercontinuum) of power per nanometer making them suitable sources for DOT. 
Single wavelengths can be generated using high repetition rate pulsed laser diodes providing tens to hundreds picosecond long pulses \cite{eda1999multichannel,grosenick2003,ntziachristos2000,cochran2019hybrid}. 
The main advantage of them over the Ti:Sapphire and supercontinuum lasers are lower cost, easier operation and smaller footprint in the expense of single wavelength operation and lower output power.}
Measurements and light detection in TD-DOT have been based on photon counting methods  and/or time-gated detectors \cite{eda1999multichannel,schmidt200032,poulet2003comparisonTR,selb2006time,farina2017multiple,milej2014time,cooper2014monstir,pifferi2016new,cochran2019hybrid}.
These are known to be sensitive, have very high dynamic range, and they are relative fast methods for detecting light.
The sensitivity and high dynamic range enable measurements both near the light sources and over relatively large distances of diffuse medium. 
Furthermore, the fast repetition rate of lasers enables detection of the entire TPSF in a reasonable time. 
However, the photon counting methods require a large number of repetitions of light illuminations in order to capture the entire TPSF.
TD-DOT has been applied, for example, in imaging tissue-mimicking phantoms \cite{schmidt200032,eda1999multichannel,farina2017multiple} and breast imaging \cite{enfield2011monitoring,cochran2019hybrid}.
Further, multi-channel NIRS has been applied for example in brain studies \cite{selb2006time,milej2014time}.

{Usage of picosecond pulses with photon counting detection in TD-DOT has been motivated by good temporal resolution, and the high repetition rate by fast acquisition time for  \emph{in vivo} imaging applications \cite{pifferi2016new}.
It has also been anticipated that a high temporal resolution can provide a good spatial resolution, especially if early photons are detected \cite{ntziachristos2001, turner2007inversion, valim2012effect}.}
On the other hand, it has been recently shown that, in diffuse regime, information content of TD-DOT data is on low frequencies, such that the reconstructed images are nevertheless low resolution \cite{mozumder2020time}.
That is, in diffuse medium, i.e. in a highly scattering medium of size larger than several mean scattering lengths, only few frequencies are required to reconstruct absorption and scattering with same resolution as with a full TD-DOT data \cite{mozumder2020time}. 

In this paper, we study feasibility of nanosecond light sources in TD-DOT.
Our work is motivated by the aspiration to enhance and develop DOT systems that could be implemented together with other imaging modalities.
Such multimodality approaches include, for example, combining DOT with EEG, MRI and ultrasound  \cite{pifferi2016new}, and photoacoustic tomography \cite{lavaud2020,xu2013,yang2013}.
Nanosecond lasers have been previously utilised in few NIRS studies to examine, for example, fruits \cite{tsuchikawa2002,kurata2008}, optical phantoms \cite{esmondewhite2009} and brain function \cite{ando2019}. 
Further, in \cite{grosenick2003,zhao2011} pulsed laser diodes with a pulse width between $100-400 \, {\rm ps}$ were utilised in breast and phantom imaging, respectively.
However, to our knowledge, nanosecond lasers have not been utilised in tomography similarly as picosecond lasers are used in the state-of-the-art TD-DOT systems.
If light illumination in DOT is based on nanosecond sources, measurements can be performed using relatively robust measurement setups with standard  analog-to-digital converter waveform digitizers, such as digital oscilloscopes.
In that case, the full TPSF can be measured directly without a need for time-gating or photon counting.

In this paper, feasibility of nanosecond lasers is studied with numerical simulations  to evaluate different aspects and limitations of such systems in TD-DOT. 
These aspects include amplitude spectrum (frequency content) of data, coarse temporal sampling and slow response time of detection, and light source variations. 
The results are compared against TD-DOT simulations with picosecond light sources.
Furthermore, feasibility of an experimental TD-DOT system with a nanosecond  Nd:YAG laser source is evaluated with phantom measurements.
The same laser has previously been utilised in photoacoustic tomography experiments \cite{sahlstrom2021}. 

The rest of the paper is organised as follows.
Theory and models for TD-DOT are described in Section \ref{sec:theory} and the experimental setup is described in Section \ref{sec:materials}. 
The numerical simulations are reported in Section \ref{sec:simu}, followed by experiments in Section \ref{sec:expt}. 
Finally, the results are discussed and concluded in Section \ref{sec:conclusions}. 


\section{Theory}
\label{sec:theory}

In DOT measurements, visible or near-infrared  light is introduced to the imaged target, and the amount of transmitted light is measured on various positions on the boundary using light sensitive detectors. 
This measurement is then repeated for multiple illumination positions leading a set of DOT data. 
Then, an image of the optical parameters is reconstructed from the measured data.
In this work, we consider estimation of absorption  and (reduced) scattering.

\subsection{Modelling light propagation}

Let us consider domain $\Omega \subset \R ^d$ with boundary  $\partial \Omega$ where $d$ is the dimension of the domain ($d=2 \, \rm{or} \, 3$).
Propagation of light in a diffuse medium can be described using the  diffusion approximation (DA) to the radiative transfer equation \cite{Ishimaru, arridge1999optical}.
The time-domain DA together with a Robin boundary condition is
\begin{equation}
\label{eq:da1}
   \left(-\nabla \cdot \frac{1}{{d}(\mua(r)+\mus(r))}
 \nabla + \mua(r) + \frac{1}{c}\frac{\partial}{\partial t} \right) \Phi (r,t)  = 0, \, r \in \Omega 
\end{equation}
\begin{equation}
\label{eq:da2}
   \Phi(r,t)+\frac{1}{2 \gamma_d} \frac{1}{d(\mua (r)+\mus(r))} \alpha \frac{\partial \Phi(r,t)}{\partial \hat{n}} = \left\lbrace \begin{array}{l} 
         \frac{Q(r,t)}{\gamma_d},  \, r \in s \\
         0, \, r \in \partial \Omega \setminus s \end{array} \right. 
\end{equation}
where  $\Phi(r,t)$ is the photon density at a point $r$ and time instance $t$, $\mua(r)$ is the absorption coefficient, and $\mus(r)$ is the reduced scattering coefficient, $c$ is the speed of light in the medium, and  $Q(r,t)$ is the pulsed (temporal) light source at source positions $s$ \cite{arridge1999optical}. 
Further,  parameter $\gamma_d$ is a dimension dependent constant that takes values $ \gamma_2$ = $1/\pi$ and $ \gamma_3$ = $1/4$, $\alpha$ is a parameter governing reflection at the boundary $\partial \Omega$, and $\hat{n}$ is an outward unit vector normal to the boundary.

The measurement data in DOT is the boundary exitance $\Gamma^+(t,r),\; t = 1,...,T$, where $T$ is the temporal range of the output signal at detector positions $r\subset \partial\Omega$. 
The exitance can be solved from photon density as 
\begin{equation}
\label{eq:extitance}
\Gamma^+(r,t) = -\frac{1}{{d}(\mua(r)+\mus(r))} \frac{\partial \Phi(r,t)}{\partial \hat{n}} = \frac{2\gamma_d}{\alpha}\Phi(r,t).
\end{equation}

Fourier transform of the time-domain DA
(\ref{eq:da1})-(\ref{eq:da2}), results in the frequency domain DA 
\begin{equation}
\label{eq:da_w1}
  \left(-\nabla \cdot \frac{1}{{d}(\mua(r)+\mus(r))}
 \nabla + \mua(r) + \frac{\rmi \omega}{c} \right) \Phi (r,\omega)= 0, \, r \in \Omega 
\end{equation}
\begin{equation}
\label{eq:da_w2}
 \Phi(r,\omega)+\frac{1}{2 \gamma_d} \frac{1}{d(\mua (r)+\mus(r))} \alpha \frac{\partial \Phi(r,\omega)}{\partial \hat{n}} = \left\{\begin{array}{l} 
         \frac{Q(r,\omega)}{\gamma_d},  \, r \in s \\
         0, \, r \in \partial \Omega \setminus s \end{array} \right. 
\end{equation}
where $\Phi(r,\omega)$ is the photon density and $Q(r,\omega)$ is the light source modulated at an angular frequency $\omega$.
Furthermore, the frequency domain exitance can be solved from the photon density as 
\begin{equation}
\label{eq:extitance_w}
\Gamma^+(r,\omega) = -\frac{1}{{d}(\mua(r)+\mus(r))} \frac{\partial \Phi(r,\omega)}{\partial \hat{n}} = \frac{2\gamma_d}{\alpha}\Phi(r,\omega).
\end{equation}

Utilising the relation between the time domain and frequency domain light transport models, it is possible to convert the measured TD-DOT data to frequency domain, and to use that data at one or several frequencies \cite{mozumder2020time}.
For light sources with a finite temporal length, the measurable data $\Gamma^+(r,t)$ can be expressed as a convolution ($\ast$) of the source $Q(r,t)$ and exitance due to delta source $\Gamma^{+}_{\delta}(r,t)$ as
\begin{equation}
    \label{eq:source_convolution}
    \Gamma^+(r,t) = \Gamma^{+}_{\delta}(r,t) \ast Q(r,t)
\end{equation}
implying that taking a Fourier transform results in product ($\cdot$) of their Fourier transforms (by convolution theorem)
\begin{equation}
    \label{eq:source_convolution_freq}
    \mathcal{F}\Gamma^+(r,t) = \mathcal{F}\Gamma^{+}_{\delta}(r,t) \cdot \mathcal{F}Q(r,t)
\end{equation}
Thus, for frequency domain computations, the Fourier transformed time-domain data needs to be divided by the Fourier transform of the source term, and then light transport can be modelled directly in frequency domain using (\ref{eq:da_w1})--(\ref{eq:extitance_w}) \cite{mozumder2020time}.

In this work, the solution of the DA is approximated with a finite element method \cite{arridge1993}. 
The FE-approximation of the time-domain DA (\ref{eq:da1})-((\ref{eq:da2})) is implemented as described in \cite{mozumder2020time} with the time-stepping implemented with a Crank–Nicholson scheme.
Further, the frequency domain FE-computations are implemented and Fourier series approximation is utilised similarly as described in \cite{mozumder2020time,mozumder2020evaluation}.
The TOAST++ software \cite{schweiger2014toast++} is utilised in the FE-integrations.  

\subsection{Image reconstruction}

Let us denote the vectors of unknown absorption and (reduced) scattering coefficients as $\mua = \left( \mu_{\rm{a},1}, \, \ldots , \mu_{{\rm a},N} \right)^{\rm T} \in \R ^N$ and $\mus = \left( \mu'_{{\rm s},1}, \, \ldots , \mu'_{{\rm s},N}  \right)^{\rm T} \in \R ^N$  where $N$ is the size of the discretisation.
Further, let the measurement data vector be $\Gamma_{\rm meas} \in \R^{N_m}$ where $N_m$ is the number of discretised data points.
In TD-DOT, the data is the measured TPSF. 
In the FD-DOT, data typically is the logarithm of amplitude and phase of  (complex) exitance. 
Further, let us denote the solution of the forward model, that maps the absorption and scattering parameters to the data, as  $\Gamma^+(\mua,\mus)$.

In DOT, typically two types of images are reconstructed: absolute and difference images. 
In absolute imaging, one aims at estimating absolute values of optical parameters using a single set of measurements during which the target is assumed to be non-varying.
In difference imaging, one is interested in the change in optical parameters between two measurements. 

Let us first consider  absolute imaging.
The image reconstruction problem can be written as a minimisation problem 
\begin{equation}
    \label{eq:map_abs}
 (\hat{\mu}_{\rm a},\hat{\mu}'_{\rm s})  =  {\arg \: \min}_{\mua,\mus} \left\lbrace \frac{1}{2} \left\Vert L_{e}(\Gamma_{\rm meas}-\Gamma^{+}(\mua,\mus)) \right\Vert^2 + \frac{1}{2} \left\Vert L_{\mu_{\rm a}}(\mua-\eta_{\mu_{\rm a}}) \right\Vert^2 + \frac{1}{2} \left\Vert L_{\mu'_{\rm s}}(\mus-\eta_{\mu'_{\rm s}})\right\Vert^2 \right\rbrace
\end{equation}
where 
$L_e$ is a weighting matrix that, from the statistical point of view, can be interpreted as the Cholesky decompostion of the inverse of the noise covariance matrix, i.e. $\Gamma_e^{-1}=L_e^{\rm T}L_e$ \cite{Kaipio,tarvainen2010}. 
Further, the two latter terms in the minimised functional (\ref{eq:map_abs}) present prior information of the target, where  $\eta_{\mu_{\rm a}}$  and $\eta_{\mu'_{\rm s}}$ are the means and $ L_{\mu_{\rm a}}$ and $L_{\mu'_{\rm s}}$ are the Cholesky decompostion of the covariance matrices of the prior model for absorption  $\Gamma_{\mu_{\rm a}}^{-1}=L_{\mu_{\rm a}}^{\rm T}L_{\mu_{\rm a}}$ and scattering $\Gamma_{\mu_{\rm s}}^{-1}=L_{\mu_{\rm s}}^{\rm T}L_{\mu_{\rm s}}$. 
In this work, the minimisation problem (\ref{eq:map_abs}) is solved using a Gauss-Newton method \cite{Schweiger2005,tarvainen2008}.

In difference imaging, two measurements are performed and the aim is to reconstruct the change in the optical parameters between these measurements \cite{arridge1999optical}. 
Consider data $\Gamma_{{\rm meas}}^1$ and $\Gamma_{{\rm meas}}^2$ of two measurements obtained from a target with optical parameters ($\mua^1,\mus^1$) and ($\mua^2,\mus^2$), respectively. 
The aim in difference imaging is to reconstruct the change in the optical parameters $\delta \mua = \mua^2-\mua^1$ and $\delta \mus=\mus^2-\mus^1$ based on the difference of the measurements as
\begin{equation}\label{modeldiff}
     \delta \Gamma_{\rm meas} = \Gamma_{{\rm meas}}^2-\Gamma_{{\rm meas}}^1 = J (\mua^1,\mus^1)\begin{pmatrix} \delta \mua \delta \mus \end{pmatrix} 
    \end{equation}
where the  $J$ is the Jacobian of the forward model $\Gamma^+$
 evaluated at $(\mua^1,\mus^1)$ \cite{mozumder2020evaluation}.  
The change in optical parameters is estimated by solving a minimisation problem
\begin{equation}
    \label{eq:map_diff}
 (\delta\hat{\mu}_{\rm a},\delta\hat{\mu}'_{\rm s})  =  {\arg \, \min}_{\delta\mua,\delta\mus} \left\lbrace \frac{1}{2} \left\Vert L_{\delta e}\left(\delta \Gamma_{\rm meas}-J(\mua^1,\mus^1)\begin{pmatrix} \delta \mua \delta \mus \end{pmatrix} \right) \right\Vert^2 + \frac{1}{2} \left\Vert L_{\delta\mu_{\rm a}}\delta\mua \right\Vert^2 + \frac{1}{2} \left\Vert L_{\delta\mu'_{\rm s}}\delta\mus\right\Vert^2 \right\rbrace
\end{equation}
where $L_{\delta e}, L_{\delta \mu_{\rm a}}$ and $L_{\delta \mu'_{\rm s}}$ are the Cholesky decompositions of the inverse of the covariance matrices of the noise and prior model for the optical parameter change \cite{mozumder2020evaluation}. 
{In this work, the difference imaging minimisation problem (\ref{eq:map_diff}), is solved using a MATLAB built-in \texttt{mldivide} function.}

In this work, we use Gaussian Ornstein-Uhlenbeck prrocess \cite{rasmussen2006} as the prior model for absorption and scattering. 
The Onrnstein-Uhlenbeck covariance function is of the form
\begin{equation}
\label{eq:prior_covariance}
\Gamma_{\mu} = \sigma^2 \Xi
\end{equation}
where $\sigma$ is the standard deviation of the prior and $\Xi$ is a matrix which  has its elements defined as
\begin{equation}
\label{eq:ornstein_uhlenbeck_covariance}
  \Xi_{ij} = \exp( - || r_i - r_j ||  / \ell )
\end{equation}
where  $i$ and $j$ denote the row and column indices of the matrix, $r_i$ and $r_j$ denote the positions of the discretisation points, and $\ell$ is the characteristic length scale of the prior describing the spatial distance that the parameter is expected to have (significant) spatial correlation for \cite{pulkkinen2014,mozumder2020evaluation}.


\section{Materials and methods}
\label{sec:materials}
\subsection{DOT system}

\begin{figure}[tbp!]
\begin{center}
\begin{tabular}{c}
\includegraphics[height=9.5cm]{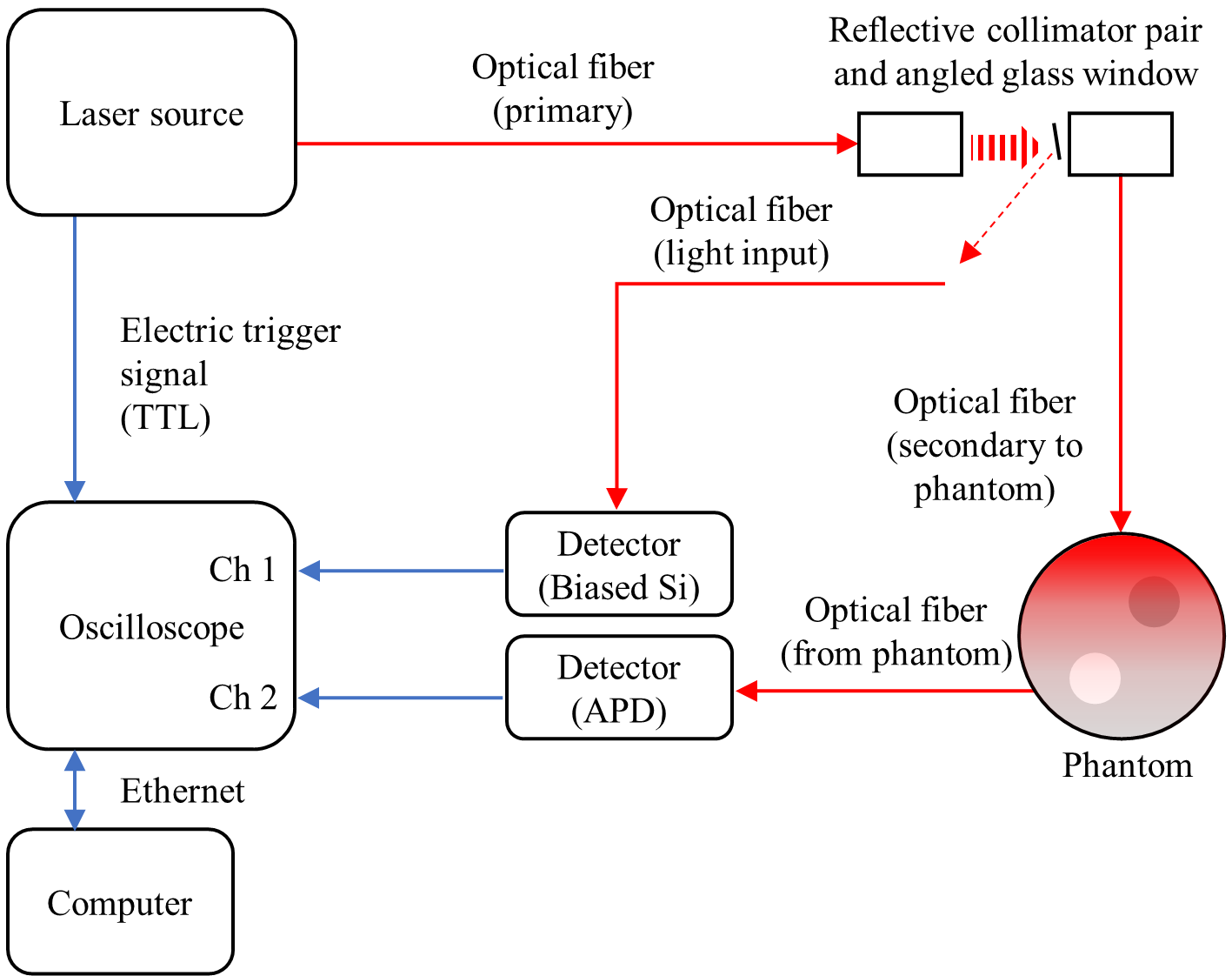}
\end{tabular}
\end{center}
\caption 
{ \label{fig:Setup}
Illustration of the time-domain diffuse optical tomography system used in the study.} 
\end{figure}

The DOT system utilised in this work is illustrated in Fig. \ref{fig:Setup}.
In the system, light source is a nanosecond Nd:YAG laser (neodymium-doped yttrium aluminum garnet) combined with optical parametric oscillator (model NT352B; Ekspla Uab, Lithuania). 
The laser can operate at 670 to 2600 nm wavelength band. The pulse energy at 670 to 825 nm band is $>$ 90 mJ and pulse repetition rate is 10 Hz. The pulse duration was approximately $3 \, {\rm ns}$. The laser was operated at 700 nm wavelength in the experiments. 
The laser pulse energy was $0.85 \, {\rm mJ}$ (average energy during the measurement session; $\pm 0.02 \, {\rm mJ}$ standard deviation when averaged over 100 pulses) and was measured with optical power meter with pyroelectric detector (models StarBright and PE50BF-C, respectively; Ophir Photonics, Israel). This value was measured from the fiber end which was connected to the target.

The laser output was directed to $3 \, {\rm m}$ long main optical fiber (multimode, diameter $1 \, {\rm mm}$, numerical aperture (NA) 0.22; Ceramoptec, Germany). 
The light from this primary fiber was further directed to $2 \, {\rm m}$ long secondary fiber (multimode, diameter $1.5 \, {\rm mm}$, NA 0.39; model M134L02, Thorlabs) using two reflecting collimators (models RPC12FC-P01 and RPC12SMA-P01; Thorlabs)
that was connected to a measurement tank containing a liquid optical phantom.
The space between the two reflective collimators served as a point to estimate the input light profile and pulse-to-pulse variation. 
For that purpose, laser reflection from angled glass placed between the collimators was directed to an optical fiber connected to a fast ($2 \, {\rm GHz}$) biased silicon photodetector (model DET025AFC/M; Thorlabs). 
Neutral density filter in front of detector was applied to limit light level below the saturation level of the detector.

Light transmitted through the target was gathered by an optical fiber (diameter $0.6 \, {\rm mm}$, NA 0.5; model M143L02, Thorlabs) and then focused on the centre of an amplified avalanche photodetector (model APD430A/M; Thorlabs). 
The detector has temperature-compensated reverse voltage regulation and wavelength operating range from $400 \, {\rm nm}$ to $1000 \, {\rm nm}$. 
The detection bandwidth ranges from DC to $400 \, {\rm MHz}$ ($-3 \, {\rm dB}$) and rise time is approximately $1 \, {\rm ns}$. 
The detector gain was set $10$, which is the minimum value.

High-bandwidth oscilloscope (model WavePro 254HD; Teledyne LeCroy, NY, USA) measured the signals of both optical detectors. 
The oscilloscope has measurement bandwidth up to $2.5 \, {\rm GHz}$, 12-bit dynamic range, and it was operated in $50 \, {\Omega}$ termination and coaxial cables.  
The sampling period was set to $50 \, {\rm ps}$. 
Signal acquisition mode was a sequence mode which enables high accuracy measurements when the pulse repetition rate is low. 

Each measurement consisted of $200$ pulses. 
Oscilloscope was triggered using transistor-transistor logic (TTL) signal with constant well-defined shape and amplitude from the laser Q-switch. 
TTL signal was used as the trigger instead of light reflection from the glass plate since the beam profiles of laser pulses detected by the biased photodetector were found to be varying and uneven in amplitude and shape. 
Trigger offset variation  was found to be less than $50 \, {\rm ps}$ indicating good temporal repeatability.
Time from the electrical trigger to laser pulse generation also varied. According to the laser manufacturer, standard deviation of the jitter between the trigger and laser pulse is approximately $0.2 \, {\rm ns}$. 
In this work, we measured laser pulses and studied effect of their variations by simulations. 

The signal detection under computer guidance was operated using LabView software (National Instruments, TX, USA) through an ethernet connection. 
After each measurement the recorded $200$ source pulses and $200$ measurement signals were averaged to provide time-domain source signal and measurement data.

\subsection{Optical phantom}
\label{sec:phant}

Optical phantom used in the study was a liquid phantom inside a black plastic cylindrical tank. 
The inner height, inner diameter, and wall thickness of the tank were $155 \,  {\rm mm}$, $80 \, {\rm mm}$ and $5 \, {\rm mm}$, respectively. 
On the tank wall, at the height of $77.5 \, {\rm mm}$,  16 holes were drilled using $22.5^{\circ}$ angular spacing. 
Eight of them had diameter of $3.2 \, {\rm mm}$ and were used only as light detection points. 
Another eight holes had $10 \, {\rm mm}$ diameter and were used both light source and light detection points. 
Liquid and the fibres were separated by a plastic membrane window (thickness $20 \, \mu \textrm{m}$, transmission approximately $95 \, \%$ at $700 \, {\rm nm}$).
The detector fibre having standard SM905 connector was directly connected and fixed to the detection points having $3.2 \, {\rm mm}$ hole. 
When the detection was made from $10 \, {\rm mm}$ holes, custom-made steel adapter and SMA bulkhead adapter (HASMA; Thorlabs) were used to accommodate the fiber head to hole. 
Similar-type adapter was used when the source fibre was connected to 10 mm holes. 
Adapter positioned the secondary source fiber end at 10 mm distance from the plastic membrane and provided 8 mm diameter exit for light beam. The 
estimated fluence on the phantom surface was $1.7 \, {\rm mJ/cm^2}$.

Clear glass tubes (Duran®; Duran Wheaton Kimble, NJ, USA) were used as a container for the liquid optical inclusions that provided different optical properties compared to the background liquid. 
The glass tubes had $130 \, {\rm mm}$ height, $13.5 \, {\rm mm}$ outer diameter, and $1 \, {\rm mm}$ wall thickness. 
The tank had a removable plastic cover and the tubes were fixed to the cover which kept them straight and in correct position inside the liquid. 
The tank was filled with the background liquid up to around $130 \, {\rm mm}$ height during the measurements.

Phantom liquid was made of degassed deionized water, Intralipid ($20 \, \%$, Fresenius Kabi, Sweden) and India Ink (Royal Talens, the Netherlands). The background solution had $1 \, \%$ concentration of Intralipid resulting in reduced scattering $\mu_{\rm{s}}' = 1 \, \textrm{mm}^{-1}$ \cite{DiNinni2012,Grabtchak2012}. 
Absorption at $700 \, {\rm nm}$ due to the Intralipid and water was estimated to be $\mua = 0.0008 \, \rm{mm}^{-1}$ \cite{Grabtchak2012}. 
This was tuned with India ink to get higher absorption \cite{DiNinni2010}. 
Ink was diluted in degassed deionized water and sonicated to provide absorption $\mua = 0.0056 \, \textrm{mm}^{-1}$ providing total absorption  $\mua = 0.0064 \, \rm{mm}^{-1}$ for the background. 
Scattering inclusion was made by increasing the amount of Intralipid to $5 \, \%$ in background liquid and assuming linear dependence between the scattering and Intralipid concentration. 
The estimated reduced scattering coefficient was therefore $\mus = 5 \, \textrm{mm}^{-1}$. 
The absorbing inclusion was made by increasing the concentration of India ink to give absorption $\mua = 0.032 \, \textrm{mm}^{-1}$.

\subsection{Tomographic measurement}

\begin{figure}[tbp!]
\begin{center}
\includegraphics[width=6.2cm]{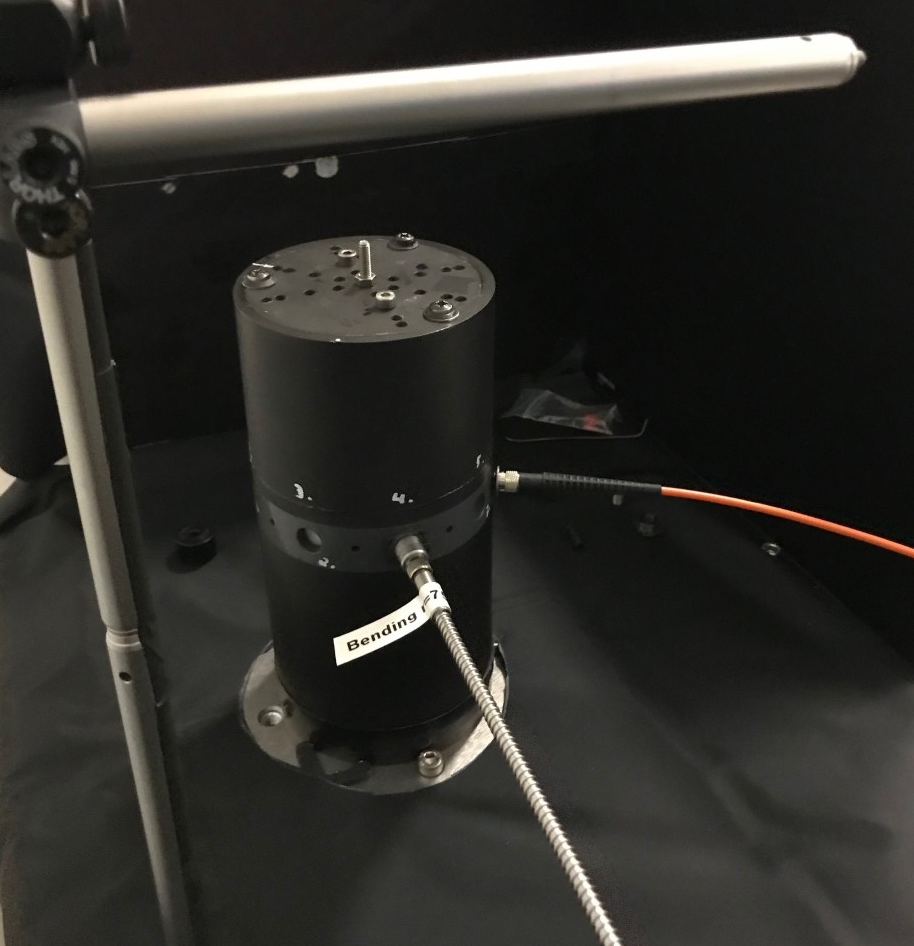}
\includegraphics[width=7.5cm]{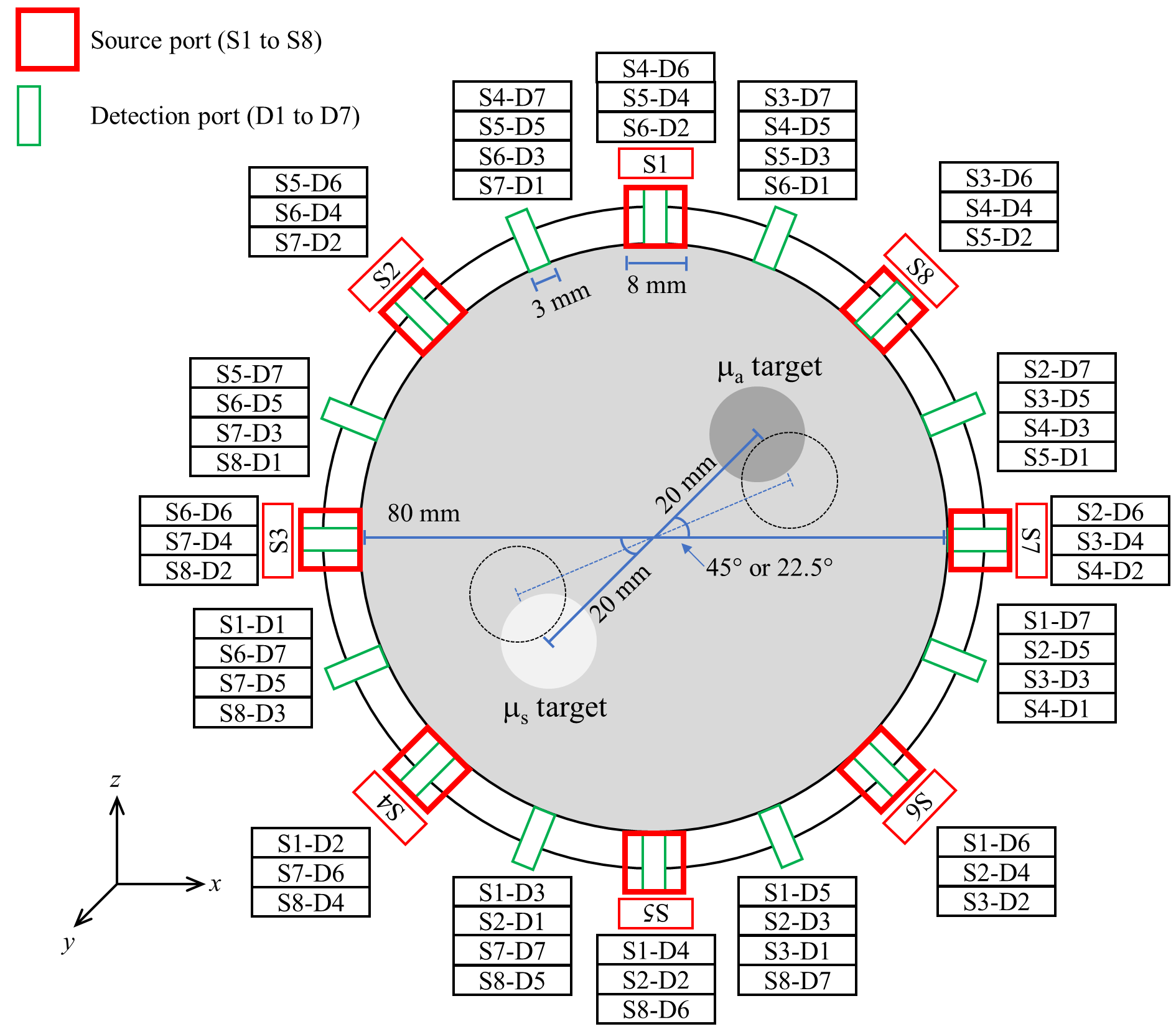}
\end{center}
\caption 
{ \label{fig:MeasProtocboth}
Experimental phantom tank (left) and illustration of source and detector locations, measurement protocol and inclusion positions in the tomography experiments (right).} 
\end{figure}

Tomographic measurement sets were made using manual source and detector fiber positioning. 
Three measurements were done: one with the liquid phantom (background measurements) and two having both scattering and absorbing inclusion tubes placed in the liquid at two different orientations.  
Each measurement set was done using eight different source locations and seven detector locations per source positions. 
Therefore, the total number of measurements was $56$ per tomography.
The source and detector positions and inclusion orientations are illustrated in Fig. \ref{fig:MeasProtocboth}.

\section{Simulations}
\label{sec:simu}

Numerical simulations were carried out to study the following aspects related to TD-DOT with nanosecond light sources.
Firstly, we wanted to study, if data obtained with nanosecond sources is lower in information such that it would lead to different quality reconstructions than those obtained using picosecond light sources.
Therefore, frequency spectra of data simulated by nanosecond and picosecond sources were compared. 
Further, a comparison between absolute and difference image reconstruction in DOT using nanosecond and picosecond light sources  and one or multiple frequencies was performed.
Secondly, if DOT measurements are performed using  digital oscilloscopes, temporal sampling is coarser and response time of detection slower when compared to data given by time-correlated photon counting methods.
Therefore, we studied the effect of the temporal sampling of detection in nanosecond TD-DOT for data and image reconstruction. 
The slower response time of detection corresponds to low-pass filtering in modelling, and thus it limits the frequencies that can be utilised in image reconstruction.
Thirdly,  nanosecond lasers, such as the one used in our setup, can suffer large variations in light pulse width and shape.
Therefore, light pulse variations in nanosecond TD-DOT data were measured and their impact on reconstructions were studied. 
The results were compared against picosecond sources with less variations.
The simulations were carried out in a Fujitsu Celcius W550 desktop workstation, with Intel\textregistered Xeon(R) W-2125 CPU @ 4.00GHz$\times$8, using MATLAB (R2017b, Mathworks, Natick, MA). 

\subsection{Data simulation}

In numerical simulations, a circular domain $\Omega \subset \mathbb{R}^2$  with a radius of $25 \, {\rm mm}$ was considered. 
The setup consisted of 16 sources and 16 detectors. 
The source and detector optodes were modelled as Gaussian surface patches with $2 \, {\rm mm}$ width, located at equi-spaced angular intervals on the boundary $\partial\Omega$.
We studied a target with background optical parameters $\mu_{{\rm a}} = 0.006 \, \rm{mm}^{-1}$ and $\mu'_{{\rm s}} = 1 \, \rm{mm}^{-1}$ and with one absorbing $\mu_{{\rm a}} = 0.03 \, \rm{mm}^{-1}$ and one scattering $\mu'_{{\rm s}} = 5 \, \rm{mm}^{-1}$ inclusion. 
The chosen optical parameters roughly corresponded to those observed in biological tissues \cite{jacques2013optical}, and our optical phantom described earlier in Section \ref{sec:phant}.
For difference imaging, the reference data was simulated using constant background optical parameter $\mu_{{\rm a}} = 0.006 \, \rm{mm}^{-1}$ and $\mu'_{{\rm s}} = 1 \, \rm{mm}^{-1}$.

The TD data was simulated using FE-approximation of the DA (\ref{eq:da1})-(\ref{eq:da2}) in a mesh with 1369 nodes and 2622 triangular elements. 
The nanosecond source pulse was modelled as a Gaussian pulse with a full width at half maximum of $3 \, {\rm ns}$.
For comparison, data with a picosecond light sources was simulated. 
In that case, the pulse was modelled as a Gaussian pulse with a full width at half maximum of $3 \, {\rm ps}$.
The temporal discretisation for both nanosecond and picosecond data simulation was $1 \, {\rm ps}$, and the temporal range was specified as $10000 \, {\rm ps}$. 
The total number of the simulated time-resolved measurements was $12 800 000$ ($256$ combination of sources and detectors, $50 000$ time steps). 

In order to evaluate the data in the frequency domain, the time-domain data was Fourier transformed to frequency domain.
The source was deconvoluted from the data by dividing the frequency domain data by corresponding frequency domain source term obtained through Fourier-transform \cite{mozumder2020time}.
Random measurement noise, that was drawn from a zero-mean Gaussian distribution 
where the standard deviations 
were specified as $1\%$ of the simulated noise-free frequency domain data, was added to the data.

\subsection{Image reconstruction}

In the reconstructions, a FE-mesh with 1123 nodes and 2142 elements was used. 
The absolute (\ref{eq:map_abs}) and difference (\ref{eq:map_diff}) image reconstruction problems were solved using the Gauss-Newton method.
The measurement noise was modelled using the statistics of the simulated data, i.e. standard deviation was $1 \, \%$ of the  noise free data.
The parameters of Ornstein-Uhlenbeck prior were chosen such that, for absolute imaging, the prior means ($\eta_{\mu_{\rm a}}$, $\eta_{\mu_{\rm s}'}$) were set as the background optical parameters and the standard deviations $(\sigma_{\mu_{\rm a}},\sigma_{\mu'_{\rm s}})$ were set such that maximum target values corresponded to three standard deviations from the background. 
For difference imaging, the prior means  ($\eta_{\delta \mu_{\rm a}}$, $\eta_{\delta \mu_{\rm s}'}$) were set to zero and  the standard deviations $(\delta\sigma_{\mu_{\rm a}},\delta\sigma_{\mu'_{\rm s}})$ were set such that maximum target values corresponded to five standard deviations from the background. 
The characteristic length scale was set as $\ell = 8 \, {\rm mm}$. 

In addition to visual inspection, the accuracy of the estimates were evaluated by computing relative errors
\begin{equation}
\label{eq:error} 
{\rm  E}_{\mu} = 100 \% \cdot \frac{\| \hat{\mu}_{} - {\mu_{}}^{\rm target} \|}{\|{\mu_{}}^{\rm target}\|} \\
\end{equation}
where $\mu^{\rm target}$ are the simulated target distributions for absorption or scattering, and $\hat{\mu}$ are the estimated parameters interpolated to the simulation grid. 

\subsection{Comparison of data and reconstructions using nanosecond and picosecond sources}
\label{sec:sim1}

First, frequency content of data simulated by nanosecond and picosecond pulses and reconstructions computed from this data were investigated.
It was studied, if data obtained with nanosecond sources
has lower frequency content such that it would lead to different quality reconstructions than those obtained from data using picosecond light sources.

\begin{figure}[tbp!]
\begin{center}
\includegraphics[height=7cm]{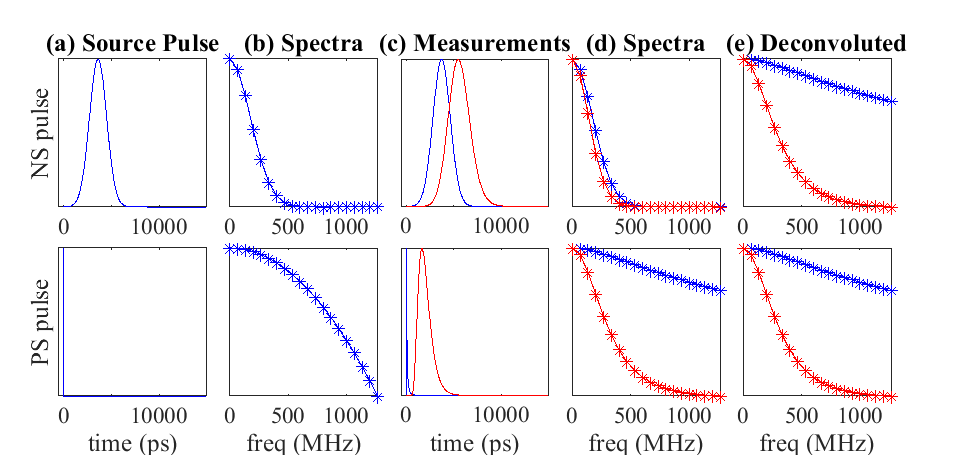}
\end{center}
\caption 
{ \label{fig:spectras} (a) Simulated source pulses of a $3 \, {\rm ns}$  source (top row) and $3 \, {\rm ps}$  source (bottom row), (b) amplitude spectra of the source pulses, (c) simulated measurement data from a near (blue) and far (red) detector using a homogeneous target, (d) amplitude spectra of the simulated data, and (e) amplitude spectra of the data deconvoluted with the source pulses.} 
\end{figure}

In order to study the frequency content of the simulated TD-DOT data using $3 \, {\rm ns}$ and $3 \, {\rm ps}$ light sources, we simulated data for detector positions located close and far from the illuminating fibre.
These detector positions correspond to locations adjacent of the source fibre and on the opposite side of the simulation domain. 
The data was simulated with a target with constant (background) optical parameters.
The simulated source pulses and the corresponding data at near and far detectors of nanosecond and picosecond sources are shown in Fig. \ref{fig:spectras} together with the amplitude spectra the signals.
The figure also shows the amplitude spectra of the simulated data deconvoluted with their source pulses.

As it can be seen from column (b) of Fig. \ref{fig:spectras}, picosecond sources have higher frequency content than nanosecond sources.
It can also be seen (columns (d) and (e)), that data measured near the source have higher frequency content than data measured far from the source.
However, although data obtained using nanosecond and picosecond sources have different spectra (column (d)), the information that they provide from the target is similar, which can be seen in the amplitude spectra of the data deconvoluted with their source pulses (column (e)).
It is especially notable that the frequency content of data measured far from the source (red lines in column (e)), that carries most information on target interior, is low. 
This predicts that resolution of images reconstructed from this data will be low. 

\begin{figure}[tbp!]
\begin{center}
\begin{tabular}{c}
\includegraphics[height=4.5cm,trim={5mm 5mm 0mm 5mm},clip]{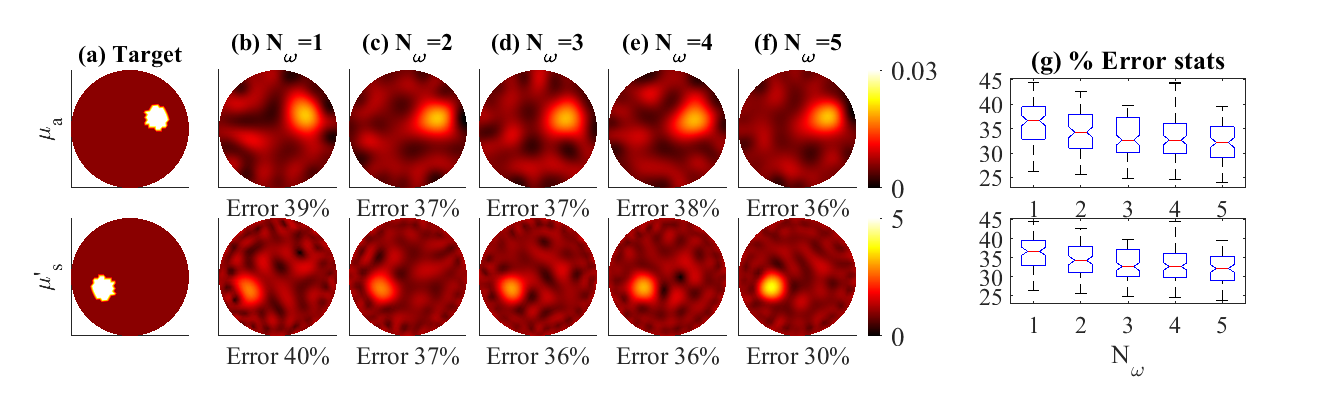}
\end{tabular}
\end{center}
\caption 
{ \label{fig:noisefree_abs}
Absolute imaging of absorption $\mua$ (top row) and scattering  $\mus$ (bottom row) distributions using data at different number of frequencies generated with a $3 \, {\rm ns}$  light source. Columns from left to right: (a) Simulated target, and (b)-(f) estimates obtained using one to five frequencies. Relative errors of the estimates (\ref{eq:error}) are give below each image. Statistics of estimation errors using 100 target distributions  are shown as `boxplots' in (g).}
\end{figure}

\begin{figure}[tbp!]
\begin{center}
\begin{tabular}{c}
\includegraphics[height=4.5cm,trim={5mm 5mm 0mm 5mm},clip]{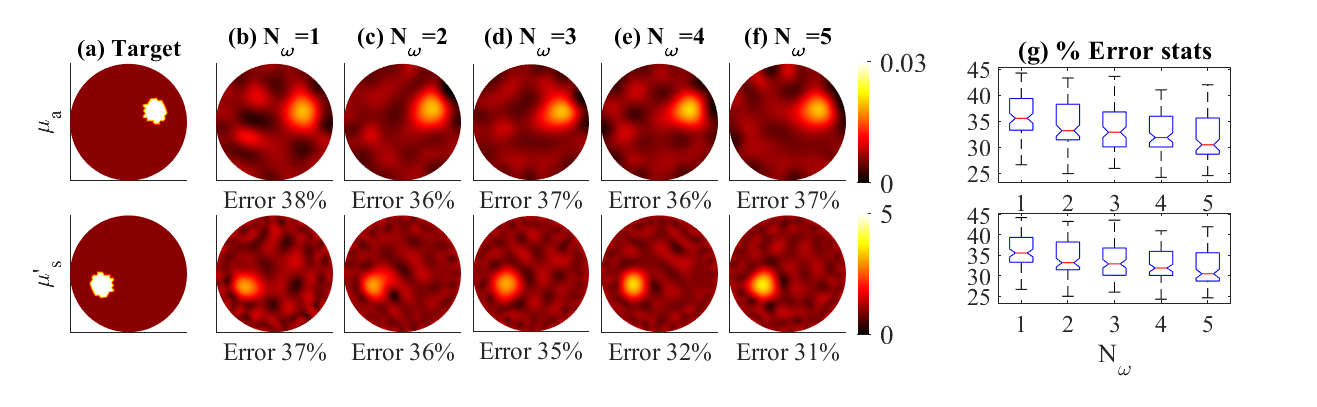}
\end{tabular}
\end{center}
\caption 
{ \label{fig:reco_abs_pico}
{Absolute imaging of absorption $\mua$ (top row) and scattering  $\mus$ (bottom row) distributions using data at different number of frequencies generated with a $3 \, {\rm ps}$  light source. Columns from left to right: (a) Simulated target, and (b)-(f) estimates obtained using one to five frequencies. Relative errors of the estimates (\ref{eq:error}) are give below each image. Statistics of estimation errors using 100 target distributions  are shown as `boxplots' in (g).}} \end{figure}

\begin{figure}[tbp!]
\begin{center}
\begin{tabular}{c}
\includegraphics[height=4.5cm,trim={5mm 5mm 0mm 5mm},clip]{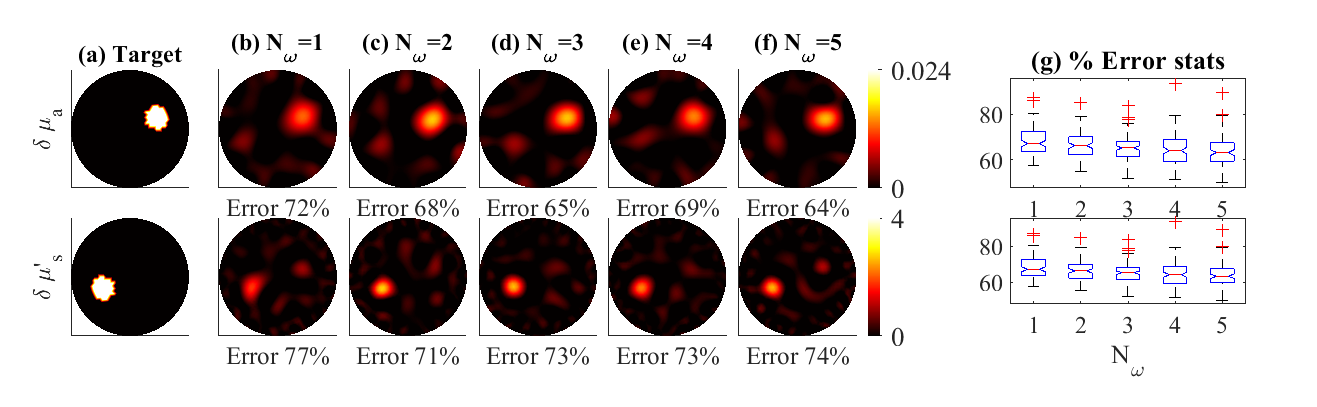}
\end{tabular}
\end{center}
\caption 
{ \label{fig:noisefree_diff}
Difference imaging of absorption $\mua$ (top row) and scattering  $\mus$ (bottom row) distributions using data at different number of frequencies generated with $3 \, {\rm ns}$  light sources. Columns from left to right: (a) Simulated target difference, and  (b)-(f) estimates obtained using one to five frequencies. Relative errors of the estimates (\ref{eq:error}) are give below each image. Statistics of estimation errors using 100 target distributions  are shown as `boxplots' in (g).} 
\end{figure}

\begin{figure}[tbp!]
\begin{center}
\begin{tabular}{c}
\includegraphics[height=4.5cm,trim={5mm 5mm 0mm 5mm},clip]{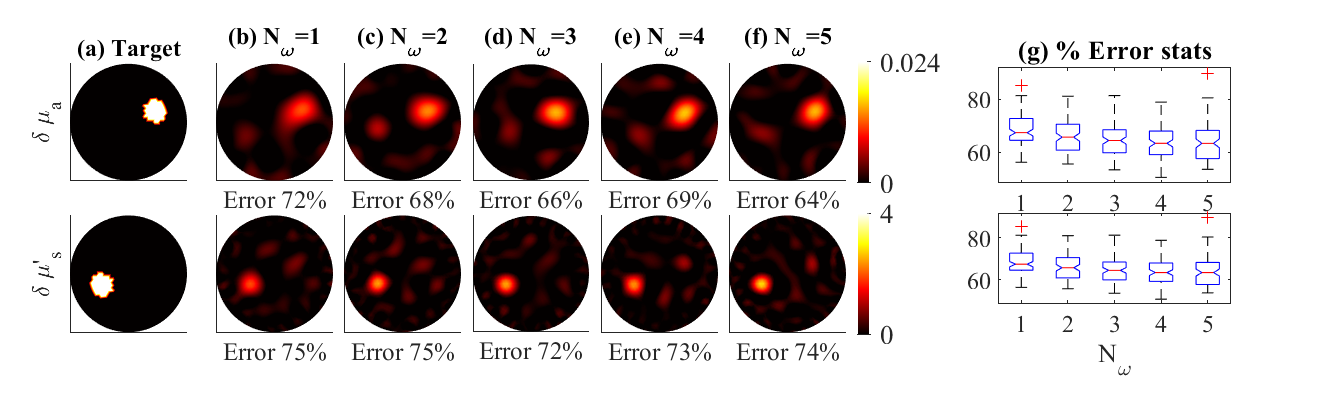}
\end{tabular}
\end{center}
\caption 
{ \label{fig:reco_diffe_pico}
{Difference imaging of absorption $\mua$ (top row) and scattering  $\mus$ (bottom row) distributions using data at different number of frequencies generated with $3 \, {\rm ps}$ light sources. Columns from left to right: (a) Simulated target difference, and (b)-(f) estimates obtained using one to five frequencies. Relative errors of the estimates (\ref{eq:error}) are give below each image. Statistics of estimation errors using 100 target distributions  are shown as `boxplots' in (g).} }
\end{figure}

The absolute and difference reconstructions using data at different number of frequencies were computed using data simulated both with $3 \, {\rm ns}$ and $3 \, {\rm ps}$  light sources. 
The frequencies that were used were $f= 66.66,133.33,200.00,266.66, 333.33 \, {\rm MHz}$. 
Notice that the first frequencies utilised are close to those typically used in frequency domain DOT systems.
The absolute reconstructions using $3 \, {\rm ns}$ light source are shown in Fig. \ref{fig:noisefree_abs} and for $3 \, {\rm ps}$  light source  in Fig. \ref{fig:reco_abs_pico} for a different number of frequencies.
Further, the difference images using $3 \, {\rm ns}$  light source are shown in Fig. \ref{fig:noisefree_diff} and for $3 \, {\rm ps}$  light source  in Fig. \ref{fig:reco_diffe_pico}. 
Statistics of relative errors of the reconstructions using 100 noise realisations are also shown. 

The results show no apparent differences between reconstructions from data generated with nanosecond or picosecond light sources.
Furthermore, both  absolute and difference imaging reconstruct the absorbing and scattering inclusion positions with similar accuracy regardless of the number of frequencies used.
On the other hand, the contrast of the images is improved if the number of frequencies is larger than one.
However, the contrast does not significantly improve if more than three frequencies are utilised.
This corresponds to our earlier findings showing that utilising multiple frequencies in imaging in diffuse medium improves reconstructions, but only up to few frequencies when compared to full time-domain data \cite{mozumder2020time}.
Based on these simulations, we chose to use data at three frequencies in later simulations of this work, as it provided reconstructions with adequate quality and accuracy.

If DOT measurements are performed using oscilloscopes, such as in our system described in Section \ref{sec:materials}, response time of detection is slow.
This can be modelled as a low-pass filter in frequency domain computations. 
In practise, this limits the frequencies in data that can be obtained and used in reconstructions. 
In the experimental system used in this paper, the cut-off frequency is approximately $400 \, {\rm MHz}$, and thus the frequencies that were used were limited below this.

\subsection{Reconstructions using data sampled at different temporal resolutions}\label{sec:sim3}

\begin{figure}[tbp!]
\begin{center}
\begin{tabular}{c}
\includegraphics[height=5cm]{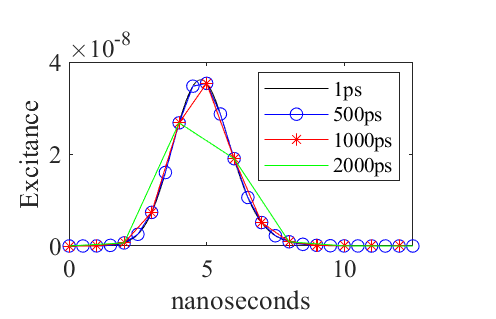}
\end{tabular}
\end{center}
\caption 
{ \label{fig:sampling}
Simulated TPSF at $1 \, {\rm ps}$ interval (black solid line) on a detector located opposite to the source position in the simulation domain. Sampled TPSFs at $500 \, {\rm ps}$, $1000 \, {\rm ps}$ and $2000 \, {\rm ps}$ temporal resolutions (blue, red and green lines, respectively). 
} 
\end{figure}

Next, the effect of low temporal sampling of data on image reconstruction was studied. 
Therefore, samples at different time-intervals were taken from the simulated time-domain data. 
The simulated data, i.e. TPSF, on a detector located opposite to the source on the other side of the simulation domain generated by a $3 \, {\rm ns}$ light pulse and the signals sampled with different temporal resolutions are illustrated in Fig. \ref{fig:sampling}.
As it can be seen, coarse sampling does not capture the original TPSF completely, that can be expected to lead artefacts in the reconstructions.

\begin{figure}[tbp!]
\begin{center}
\begin{tabular}{c}
\includegraphics[height=6cm]{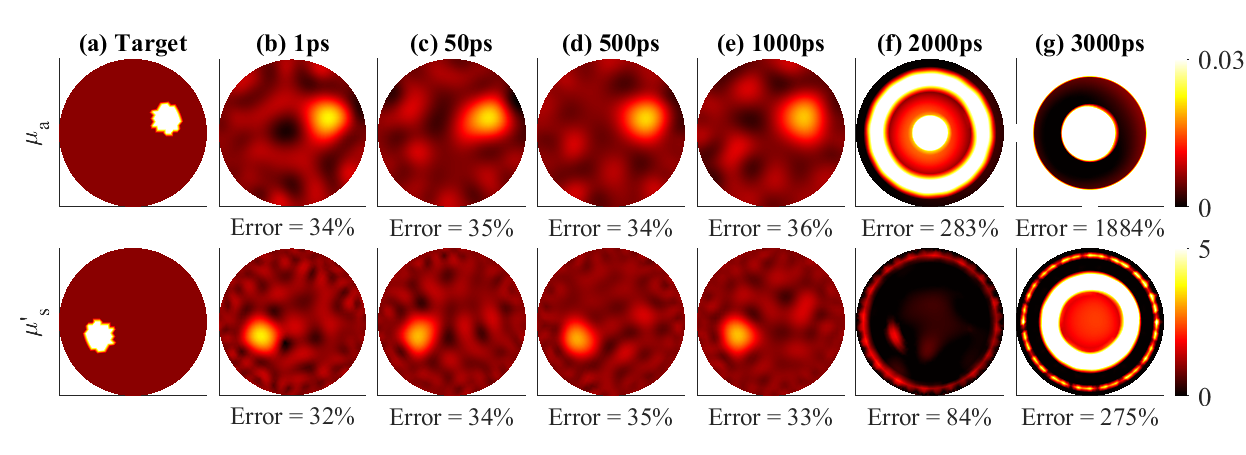}
\end{tabular}
\end{center}
\caption 
{ \label{fig:sampling_abs}
Absolute imaging  of absorption $\mua$ (top row) and scattering  $\mus$ (bottom row) distributions from $3 \, {\rm ns}$ light source data sampled at different temporal discretisations. Columns from left to right: (a) Simulated target, and estimates obtained by sampling the measured TPSFs at (b) $1 \, {\rm ps}$, (c) $50 \, {\rm ps}$, (d) $500 \, {\rm ps}$, (e) $1000 \, {\rm ps}$, (f) $2000 \, {\rm ps}$ and (g) $3000 \, {\rm ps}$ temporal resolution. Relative errors of the estimates (\ref{eq:error}) are given below each image.} 
\end{figure}

\begin{figure}[tbp!]
\begin{center}
\begin{tabular}{c}
\includegraphics[height=6cm]{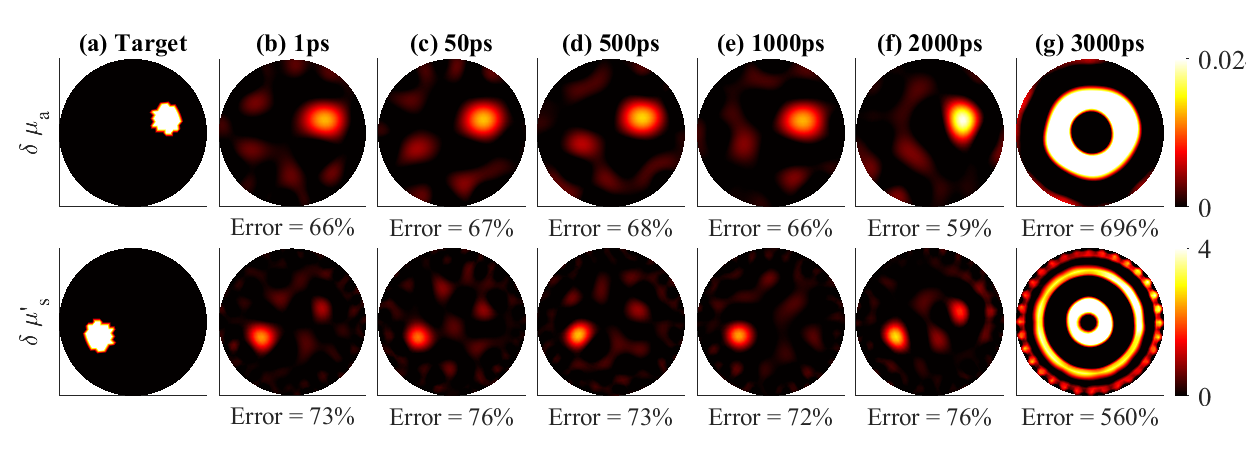}
\end{tabular}
\end{center}
\caption 
{ \label{fig:sampling_diff}
Difference imaging of absorption $\delta \mua$ (top row) and scattering $\delta \mus$ (bottom row) distributions from $3 \, {\rm ns}$ light source data sampled at different temporal discretisations. Columns from left to right: (a) Simulated target difference, and estimates obtained by sampling the measured TPSFs at (b) $1 \, {\rm ps}$, (c) $50 \, {\rm ps}$, (d) $500 \, {\rm ps}$, (e) $1000 \, {\rm ps}$, (f) $2000 \, {\rm ps}$ and (g) $3000 \, {\rm ps}$ temporal resolution. Relative errors of the estimates (\ref{eq:error}) are given below each image.} 
\end{figure}

Then, the signals with different temporal resolutions were Fourier-transformed to frequency domain where absolute and difference reconstructions were computed using three frequencies of data, $f=66.66, \, 133.33, \, 200.00 \, {\rm MHz}$.
For difference imaging, the simulated reference data was sampled using the same temporal sampling and transformed to frequency domain. 
The absolute reconstructions from data at different temporal sampling are shown in Fig. \ref{fig:sampling_abs}, and 
the difference images from data at different temporal sampling are shown in Fig. \ref{fig:sampling_diff}.
As it can be seen from Fig. \ref{fig:sampling_abs}, the absolute reconstructions obtained from data with temporal sampling between $1 \, {\rm ps}$ and $1000 \, {\rm ps}$ look qualitatively similar.
Further,  the relative errors of the estimates are approximately the same.
However, when temporal sampling decreases even more, the reconstructions suffer from artefacts, that are substantial at low temporal sampling (images with $2000 \, {\rm ps}$ and  $3000 \, {\rm ps}$ temporal resolution), and the relative errors increase.
It can further be seen from Fig. \ref{fig:sampling_diff} that difference imaging cancels out the errors caused by a low temporal sampling to some extent.
However, when the sampling is in the same level as the source pulse width, the difference imaging cannot correct image artefacts either.

In general, the simulations demonstrate that low temporal sampling should not affect on the reconstructions as long as the sampling is high enough to capture the temporal features of the data.
In our experimental setup, the temporal sampling is $50 \, {\rm ps}$ and that should not affect the reconstructions.

\subsection{Reconstructions in the presence of light source variations}\label{sec:sim2}

Nanosecond lasers suffer from larger variations in light pulse
width and shape than picosecond lasers. 
Therefore, we studied with simulations the effect of light source variations  on image reconstruction.  
The results were compared against picosecond sources with less variations.

\begin{figure}[tbp!]
\begin{center}
\begin{tabular}{c}
\includegraphics[height=6cm]{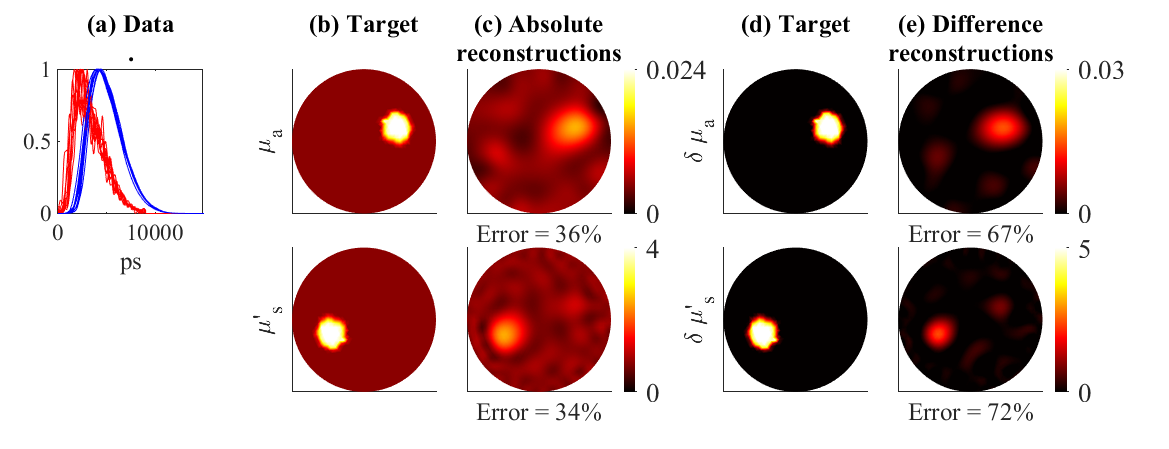}
\end{tabular}
\end{center}
\caption 
{ \label{fig:fluctuations_ns}
Effect of nanosecond source variations on absolute and difference reconstructions. (a) Normalised source pulses (red line) and the corresponding normalised simulated data (blue line). (b) Simulated target, and (c) absolute absorption $\mua$ and scattering $\mus$ reconstructions. (c) Simulated target difference, and (e) difference absorption $\delta \mua$ and scattering $\delta\mus$ reconstructions in the presence of source variations.}
\end{figure}

\begin{figure}
\begin{center}
\begin{tabular}{c}
\includegraphics[height=6cm]{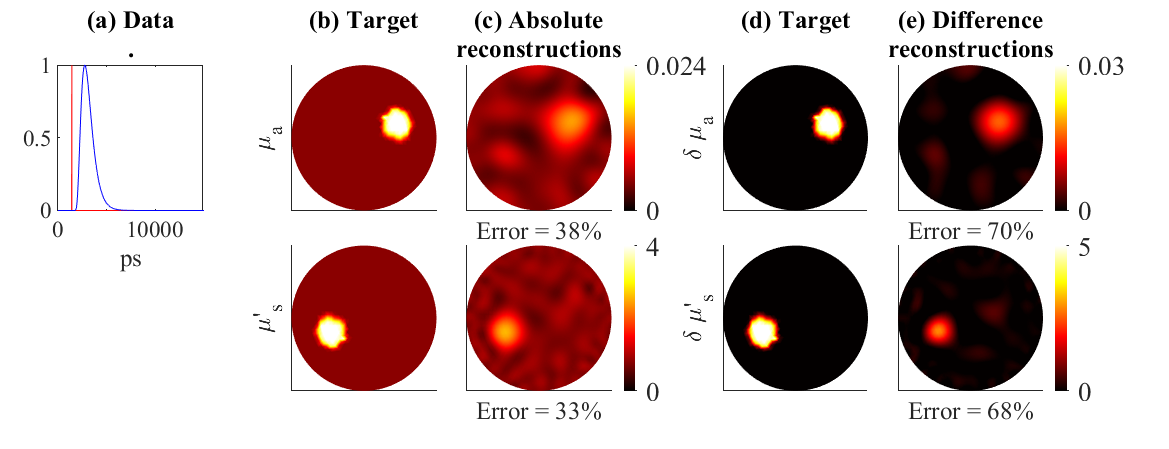}
\end{tabular}
\end{center}
\caption 
{ \label{fig:fluctuations_ps}
Effect of picosecond source variations on absolute and difference reconstructions. (a) Simulated normalised source pulses (red line) and the corresponding normalised simulated data (blue line). (b) Absolute absorption $\mua$ and scattering $\mus$ reconstructions and (c) difference absorption $\delta \mua$ and scattering $\delta\mus$ reconstructions in the presence of source variations.} 
\end{figure}

For this, we measured $10$ light source pulses of the laser of the experimental setup, with $50 \, {\rm ps}$ sampling.
Then, these source signals were interpolated to $0.5 \, {\rm ps}$ duration and were used as light sources in simulations to simulate $10$ TD-DOT data sets.
These source pulses and data were averaged similarly as in data processing of the experimental system.
The reference data for difference imaging was simulated and processed similarly.
Measured source pulses are show in  Fig. \ref{fig:fluctuations_ns} (a) together with the corresponding simulated data.
In order to compare the reconstructions to picosecond lasers that have less variations, a proportional source pulse variations were simulated to the picosecond light sources.
For that, $1400$ source pulses were measured and their amplitude and temporal (phase) variations were calculated to be 0.03 (arbitrary units) and $0.23 \, {\rm ns}$ respectively. 
These were then scaled to picosecond range 
and used to simulate $10$  picosecond sources and simulate TD-DOT data. 
Simulated picosecond source signals together with the corresponding simulated data are shown in Fig.  \ref{fig:fluctuations_ps} (a).

Then, the simulated signals were Fourier transformed to the frequency domain, and absolute and difference images were reconstructed similarly as earlier using data at three frequencies $f=66.66, \, 133.33, \,  200.00 \, {\rm MHz}$.
The absolute and difference reconstructions are shown in Fig. \ref{fig:fluctuations_ns} for data simulated with nanosecond light pulses and in Fig. \ref{fig:fluctuations_ps} for data simulated with picosecond light pulses. 
As it can be seen, the source variations do not result in any additional loss of image quality, compared to absolute and difference images obtained without source pulse fluctuations in Figs. \ref{fig:noisefree_abs} and \ref{fig:noisefree_diff}.
It can also be seen that there are no significant differences between reconstructions from nanosecond and picosecond lasers.
That is, although nanosecond lasers have larger variations, it seems that those do not affect the accuracy of the reconstructed images.


\section{Experiments}
\label{sec:expt}

DOT measurements were performed using the experimental system and phantoms described in Section \ref{sec:materials}.
Two cylindrical phantoms with an absorbing inclusion and a scattering inclusion approximately five times the value of the background parameters were studied.
The phantoms are illustrated in Fig. \ref{fig:MeasProtocboth} together with the measurement protocol. 
The measurements were performed using eight source locations (S1-S8) and seven detector locations (D1-D7).
Furthermore, measurements with a homogeneous reference phantom were made.
The reference data was utilised in difference imaging.
In addition, it was used to provide a computational calibration measurement for absolute imaging. 

\begin{figure}[tbp!]
\begin{center}
\begin{tabular}{c}
\includegraphics[height=5cm,trim={2mm 0mm 0mm 0mm},clip]{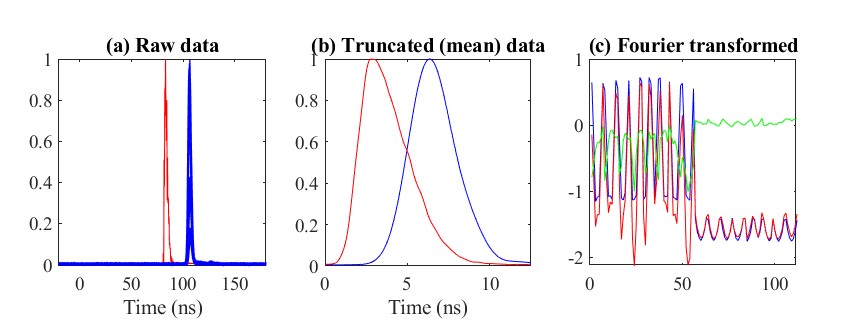}
\end{tabular}
\end{center}
\caption 
{ \label{fig:expt_data}
(a) Raw measurement data from the homogeneous reference phantom showing the source pulse (red) and the measurement pulse (blue) measured using an adjacent source-detector pair. (b) Mean data from $200$ samples of source (red) and measurement pulses (blue) in a truncated temporal interval. (c) Logarithm of amplitude (first half of the x-axis) and phase (second half of the x-axis) of the Fourier transformed measurements for all source-detector configurations from a phantom with two inclusions (red), homogeneous phantom (blue), and their difference (green).
}
\end{figure}

A raw measurement signal collected from the homogeneous reference phantom on a detector adjacent to a source is shown in Fig. \ref{fig:expt_data} (a).
The measurement data was averaged over $200$ the samples and truncated to a temporal window of the measurement pulse width as shown in Fig. \ref{fig:expt_data} (b). 
Then, the data was Fourier-transformed to the frequency domain and deconvoluted with a source by dividing the Fourier-transformed data with the corresponding Fourier-transformed source pulse.
The Fourier transformed data at frequency $f=159.36 \, {\rm MHz}$ for all source-detector combinations is shown in Fig. \ref{fig:expt_data} (c).

The difference and absolute reconstructions were computed from frequency domain data at frequency $f=159.36 \, {\rm MHz}$. 
In reconstructions, a 2D computation domain was considered. 
The domain was discretised using 1369 nodes and 2622 elements.
The absolute reconstructions were computed by minimising (\ref{eq:map_abs}) and the difference reconstructions were computed by minimising (\ref{eq:map_diff}). 
The minimisation problems were solved using Gauss-Newton method similarly as for simulated data.
For absolute imaging, the measurements with inclusions were precalibrated using the following procedure. 
Calibration coefficients for individual source-detector pairs were computed by taking the difference between the experimental reference measurements and simulated reference measurements. 
Then, this difference was subtracted from the experimental measurements with inclusions, and
the subtracted measurements were considered as calibrated data.

\begin{figure}[tbp!]
\begin{center}
\includegraphics[height=7.5cm,trim={5mm 10mm 0mm 0mm},clip]{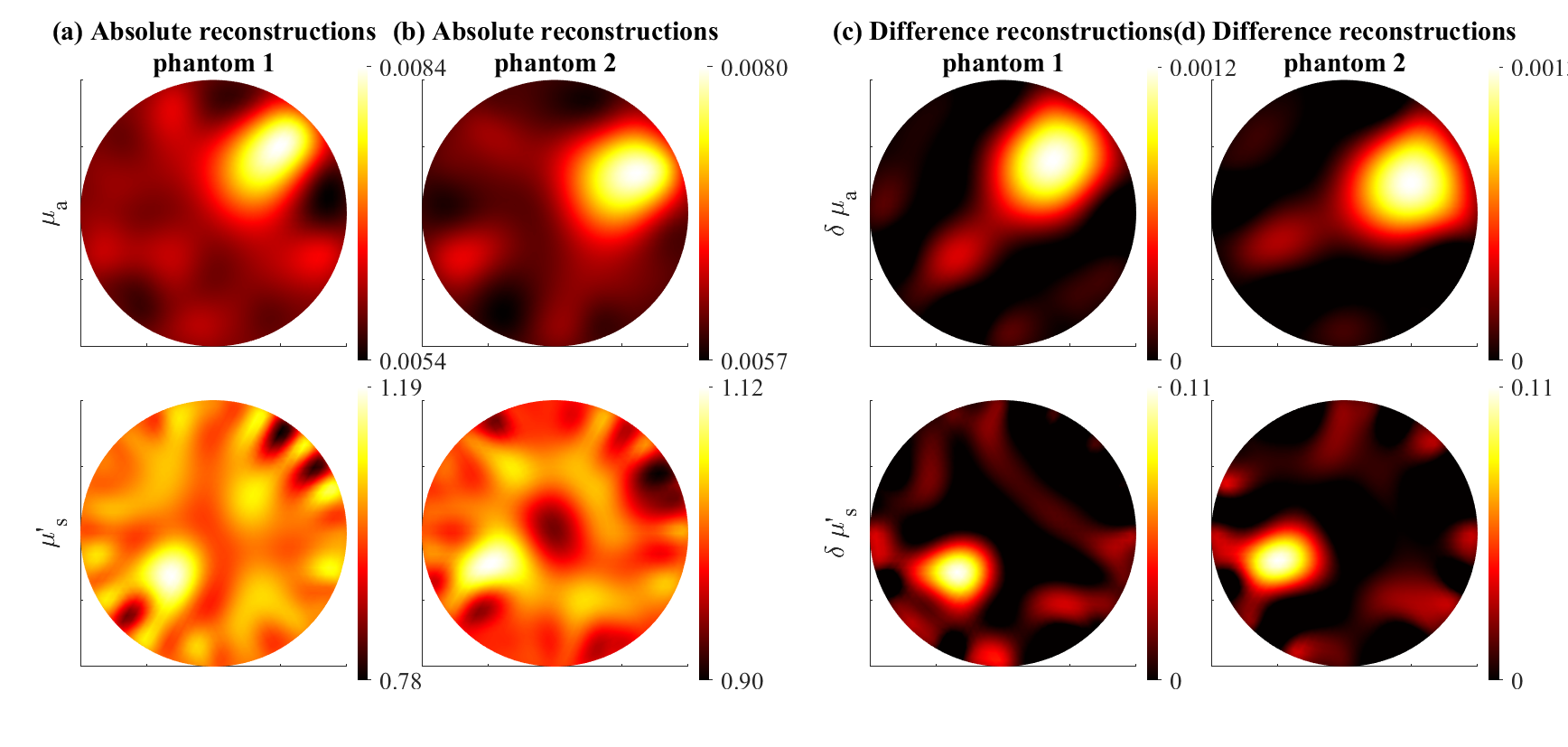}
\end{center}
\caption 
{ \label{fig:reco_expt_all}
(a) Reconstructed absolute absorption $\mua$ (first row) and scattering $\mus$ (second row) images for the two phantoms (first and second column). (b) Reconstructed difference absorption $\delta \mua$ (first row) and scattering $\delta \mus$ (second row) images for the two phantoms (third and fourth column).}
\end{figure}

Reconstructed absolute and difference images for the two phantoms are shown in Fig. \ref{fig:reco_expt_all}.
As it can be seen, the location of absorption and scattering inclusions can be distinguished both using absolute and difference imaging. 
Also the difference in inclusion positions between the two phantoms is clearly visible both in absolute and difference imaging. 
The absorption images show some cross-talk from the scattering inclusion, that is especially evident in the difference images, but the magnitude of the cross-talk is very low.
The scattering images, on the other hand, have more artefacts both in absolute and difference imaging.
The difference imaging shows less artifacts than absorption images.
That is typical in DOT since difference imaging compensates both measurement and modelling errors.
Overall, all reconstructions can be regarded as good quality DOT images. 
However, the estimated inclusion values do not reach the correct absolute values of the inclusions.
This can be due to, for example, relatively low number of measurement positions that are all located on a single plane of the 3D object.

\section{Discussion and conclusions}
\label{sec:conclusions}

In this work, feasibility of utilising nanosecond sources in TD-DOT was investigated.
Different aspects and possible limitations of TD-DOT systems with nanosecond sources were studied with simulations and experiments.

First, frequency content of time-domain DOT data simulated with nanosecond light sources was studied and compared against picosecond source systems.
The simulations verified the previous findings \cite{mozumder2020time} that information of diffuse medium is on low frequencies.
This enables image reconstruction using Fourier-transformed time-domain data using few frequencies. 
Furthermore, it enables usage of measurement electronics with slower response time.  
Second, effect of temporal sampling was studied.
The simulations showed that temporal sampling needs to be high enough to capture the TPSF.
This sampling can be achieved with standard digital oscilloscopes.
Third, nanosecond lasers can suffer from light pulse variations. 
It was shown that, although these variations are large, they do not affect significantly on image quality. 
Overall, the simulations demonstrated the capability of nanosecond sources to be utilised in TD-DOT in diffuse medium. 

Then, a prototype TD-DOT experimental system utilising a high-energy nanosecond source was constructed.
The system consisted of a nanosecond Nd:YAG laser combined with optical
parametric oscillator for light input, optical fibres and collimators for guiding the light, and avalanche photodetector and high-bandwidth oscilloscope for measurements.
The system is relatively robust.
For example, detector is temperature-compensated, it does not require cooling and can be operated in normal room lighting without saturation.
The pros of the system include broad wavelength tuning range for multiwavelength imaging as well as high energy per wavelength and pulse, that would enable directing the light to multiple fibres.
Further, relative impacts of pulse temporal spreading in long optical fibres,  inaccuracies on fibre lengths, or changes in environmental conditions such as temperature and vibrations are smaller for nanosecond pulses and could provide robustness and stability for the approach.
Furthermore, a nanosecond laser and a established digitizer based signal detection enable compatibility of the system with other techniques such as ultrasound imaging  (signal waveform detection) and photoacoustic tomography (signal waveform detection and laser source).
The cons of the system include poor laser pulse stability, that was not an issue in the experiments of the study but can be significant in some other applications, and the low pulse repetition frequency of the laser.  
Also the dynamic range and sensitivity of the system may have limitations in larger targets, which would require more research.

The DOT system was used in both absolute and difference imaging of two phantoms.
It was shown that  both absorbing and scattering objects could be reconstructed.
The locations of the inclusions were found, and the cross-talk between the absorbing and scattering targets was low.
The reconstructions could be improved, for example, by adding more measurement layers to the phantom and extending modelling to 3D.

The simulations and experiments of this work were the first study demonstrating usage of nanosecond sources in TD-DOT.
The nanosecond system can be utilised in diffuse medium, that is in highly scattering medium when the imaged target size is larger than multiple free scattering lengths.
However, its performance, for example, in dilute medium where photons travel faster from source to detector or with measurement setups with short source-detector distances were not studied.
Overall, developments of DOT systems are guided by potential applications and their different requirements on, for example, imaging depth, speed, invasiveness, scalability and multimodality \cite{pifferi2016new}. 
Therefore, further work is needed for the development of the proposed setup, and finding the applications where it could be seen as most beneficial.
The system could be further developed by building source energy specific light detection utilising, for example, neutral density filters or laser beam attenuators.
Sensitivity of the system could be increased using large diameter detection fibre bundles.

\subsection* {Acknowledgments}
This project has received funding from the European Research Council (ERC) under the European Union's Horizon 2020 research and innovation programme (grant agreement No 101001417- QUANTOM). The work has been supported by the Academy of Finland (projects 314411, 336799 Centre of Excellence in Inverse Modeling and Imaging, and 320166 the Flagship Program Photonics Research and Innovation) and Finnish Cultural Foundation (project 00200746).

\subsection* {References}

\bibliographystyle{vancouver} 
\bibliography{report}   

\begin{thebibliography}{10}

\bibitem{arridge1999optical}
Arridge SR.
\newblock Optical tomography in medical imaging.
\newblock Inverse Problems. 1999;15(2):R41.

\bibitem{gibson2005recent}
Gibson A, Hebden J, Arridge SR.
\newblock Recent advances in diffuse optical imaging.
\newblock Physics in Medicine \& Biology. 2005;50(4):R1.

\bibitem{durduran2015}
Durduran T, Choe R, Baker WB, Yodh AG.
\newblock Diffuse optics for tissue monitoring and tomography.
\newblock Rep Prog Phys. 2015;73:076701.

\bibitem{enfield2009optical}
Enfield LC, Gibson AP, Hebden JC, Douek M.
\newblock Optical tomography of breast cancer-monitoring response to primary
  medical therapy.
\newblock Targeted Oncology. 2009;4(3):219-33.

\bibitem{darne2014}
Darne C, Lu Y, Sevick-Muraca EM.
\newblock Small animal fluorescence and bioluminescence tomography: a review of
  approaches, algorithms and technology update.
\newblock Phys Med Biol. 2014;59:R1-R64.

\bibitem{grosenick2016}
Grosenick D, Rinneberg H, Cubeddu R, Taroni P.
\newblock Review of optical breast imaging and spectroscopy.
\newblock J Biomed Opt. 2016;21(10):091311.

\bibitem{hoshi2016}
Hoshi Y, Yamada Y.
\newblock Overview of diffuse optical tomography and its clinical applications.
\newblock J Biomed Opt. 2016;21(9):091312.

\bibitem{schmitz2002instrumentation}
Schmitz CH, L{\"o}cker M, Lasker JM, Hielscher AH, Barbour RL.
\newblock Instrumentation for fast functional optical tomography.
\newblock Review of Scientific Instruments. 2002;73(2):429-39.

\bibitem{arridge1998nonuniqueness}
Arridge SR, Lionheart WR.
\newblock Nonuniqueness in diffusion-based optical tomography.
\newblock Optics letters. 1998;23(11):882-4.

\bibitem{pifferi2016new}
Pifferi A, Contini D, Dalla~Mora A, Farina A, Spinelli L, Torricelli A.
\newblock New frontiers in time-domain diffuse optics, a review.
\newblock Journal of Biomedical Optics. 2016;21(9):091310.

\bibitem{selb2006time}
Selb JJ, Joseph DK, Boas DA.
\newblock Time-gated optical system for depth-resolved functional brain
  imaging.
\newblock Journal of Biomedical Optics. 2006;11(4):044008.

\bibitem{contini2015}
Contini D, Mora AD, Spinelli L, Farina A, Torricelli A, Cubeddu R, et~al.
\newblock Effects of time-gated detection in diffuse optical imaging at short
  source-detector separation.
\newblock Journal of Physics D: Applied Physics. 2015;48(4):045401 (11pp).

\bibitem{cooper2014monstir}
Cooper RJ, Magee E, Everdell N, Magazov S, Varela M, Airantzis D, et~al.
\newblock MONSTIR {II}: a 32-channel, multispectral, time-resolved optical
  tomography system for neonatal brain imaging.
\newblock Review of Scientific Instruments. 2014;85(5):053105.

\bibitem{lapointe2012}
Lapointe E, Pichette J, B{\'e}rub{\'e}-Lauzi{\`e}re Y.
\newblock A multi-view time-domain non-contact diffuse optical tomography
  scanner with dual wavelength detection for intrinsic and fluorescence small
  animal imaging.
\newblock Rev Sci Instrum. 2012;83(6):063703.

\bibitem{eda1999multichannel}
Eda H, Oda I, Ito Y, Wada Y, Oikawa Y, Tsunazawa Y, et~al.
\newblock Multichannel time-resolved optical tomographic imaging system.
\newblock Review of Scientific Instruments. 1999;70(9):3595-602.

\bibitem{grosenick2003}
Grosenick D, Moesta KT, Wabnitz H, Mucke J, Stroszczynski C, Macdonald R,
  et~al.
\newblock Time-domain optical mammography: initial clinical results on
  detection and characterization of breast tumors.
\newblock Applied Optics. 2003;42(16):3170-86.

\bibitem{ntziachristos2000}
Ntziachristos V, Yodh AG, Schnall M, Chance B.
\newblock Concurrent {MRI} and diffuse optical tomography of breast after
  indocyanine green enhancement.
\newblock PNAS. 2000;97(6):2767-72.

\bibitem{cochran2019hybrid}
Cochran JM, Busch DR, Lin L, Minkoff DL, Schweiger M, Arridge S, et~al.
\newblock Hybrid time-domain and continuous-wave diffuse optical tomography
  instrument with concurrent, clinical magnetic resonance imaging for breast
  cancer imaging.
\newblock Journal of Biomedical Optics. 2019;24(5):051409.

\bibitem{schmidt200032}
Schmidt FE, Fry ME, Hillman EM, Hebden JC, Delpy DT.
\newblock A 32-channel time-resolved instrument for medical optical tomography.
\newblock Review of Scientific Instruments. 2000;71(1):256-65.

\bibitem{poulet2003comparisonTR}
Poulet P, Zint CV, Torregrossa M, Uhring W, Cunin B.
\newblock Comparison of two time-resolved detectors for diffuse optical
  tomography: photomultiplier tube - time-correlated single photon counting and
  multi-channel streak camera.
\newblock In: Proc. SPIE 4955, Optical Tomography and Spectroscopy of Tissue V.
  vol. 4955. SPIE. International Society for Optics and Photonics; 2003. .

\bibitem{farina2017multiple}
Farina A, Betcke M, Di~Sieno L, Bassi A, Ducros N, Pifferi A, et~al.
\newblock Multiple-view diffuse optical tomography system based on time-domain
  compressive measurements.
\newblock Optics Letters. 2017;42(14):2822-5.

\bibitem{milej2014time}
Milej D, Gerega A, Kacprzak M, Sawosz P, Weigl W, Maniewski R, et~al.
\newblock Time-resolved multi-channel optical system for assessment of brain
  oxygenation and perfusion by monitoring of diffuse reflectance and
  fluorescence.
\newblock Opto-Electronics Review. 2014;22(1):55-67.

\bibitem{enfield2011monitoring}
Enfield LC, Cantanhede G, Westbroek D, Douek M, Purushotham AD, Hebden JC,
  et~al.
\newblock Monitoring the Response to Primary Medical Therapy for Breast Cancer
  Using Three-Dimensional Time-Resolved Optical Mammography.
\newblock Technology in Cancer Research and Treatment. 2011;10(6):533-47.

\bibitem{ntziachristos2001}
Ntziachristos V, Chance B.
\newblock Accuracy limits in the determination of absolute optical properties
  using time-resolved {NIR} spectroscopy.
\newblock Medical Physics. 2001;28(6):1115-24.

\bibitem{turner2007inversion}
Turner GM, Soubret A, Ntziachristos V.
\newblock Inversion with early photons.
\newblock Medical physics. 2007;34(4):1405-11.

\bibitem{valim2012effect}
Valim N, Brock J, Leeser M, Niedre M.
\newblock The effect of temporal impulse response on experimental reduction of
  photon scatter in time-resolved diffuse optical tomography.
\newblock Physics in Medicine \& Biology. 2012;58(2):335.

\bibitem{mozumder2020time}
Mozumder M, Tarvainen T.
\newblock Time-domain diffuse optical tomography utilizing truncated {F}ourier
  series approximation.
\newblock Journal of the Optical Society of America A. 2020;37(2):182-91.

\bibitem{lavaud2020}
Lavaud J, Henry M, Gayet P, Fertin A, Vollaire J, Usson Y, et~al.
\newblock Noninvasive monitoring of liver metastasis development via combined
  multispectral photoacoustic imaging and fluorescence diffuse optical
  tomography.
\newblock Journal of Biological Sciences. 2020;16(9):1616-28.

\bibitem{xu2013}
Xu C, Kumavor PD, Alqasemi US, Li H, Y~Xu SZ, Zhu Q.
\newblock Indocyanine green enhanced co-registered diffuse optical tomography
  and photoacoustic tomography.
\newblock J Biomed Opt. 2013;18(12):12600.

\bibitem{yang2013}
Yang H, Xi L, Samuelson S, Xie H, Yang L, Jiang H.
\newblock Handheld miniature probe integrating diffuse optical tomography with
  photoacoustic imaging through a {MEMS} scanning mirror.
\newblock Biomed Opt Express. 2013;4(3):427-32.

\bibitem{tsuchikawa2002}
Tsuchikawa S, Kumada S, Inoue K, Cho RK.
\newblock Application of Time-of-flight Near-infrared Spectroscopy for
  Detecting Water Core in Apples.
\newblock J Amer Soc Hort Sci. 2002;127(2):303-208.

\bibitem{kurata2008}
Kurata Y, Ikemoto Y, Tsuchikawa S.
\newblock Application of time-of-flight near infrared spectroscopy to
  fruits—permeability of pulsed laser beam into {S}atsuma mandarin, {W}hite
  grapefruit and {F}uji apples.
\newblock Journal of Near Infrared Spectroscopy. 2008;16:139-42.

\bibitem{esmondewhite2009}
Esmonde-White FWL, Burns DH.
\newblock A portable multi-wavelength near infrared photon time-of-flight
  instrument for measuring light scattering.
\newblock Journal of Near Infrared Spectroscopy. 2009;17:167–176.

\bibitem{ando2019}
Ando T, Nakamura T, Fujii T, Shiono T, Nakamura T, Suzuki M, et~al.
\newblock Non-contact acquisition of brain function using a time-extracted
  compact camera.
\newblock Scientific Reports. 2019;9:17854.

\bibitem{zhao2011}
Zhao Q, Spinelli L, Bassi A, Valentini G, Contini D, Torricelli A, et~al.
\newblock Functional tomography using a time-gated {ICCD} camera.
\newblock Biomed Opt Express. 2011;2(3):705-16.

\bibitem{sahlstrom2021}
Sahlstr{\"o}m T, Pulkkinen A, Leskinen J, Tarvainen T.
\newblock Computationally efficient forward operator for photoacoustic
  tomography based on coordinate transformations.
\newblock IEEE Transactions on Ultrasonics, Ferroelectrics, and Frequency
  Control. 2021;69(6):2172-82.

\bibitem{Ishimaru}
Ishimaru A.
\newblock Wave Propagation and Scattering in Random Media.
\newblock Academic, Newyork; 1978.

\bibitem{arridge1993}
Arridge SR, Schweiger M, Hiraoka M, Delpy DT.
\newblock A Finite Element approach to modelling photon transport in tissue.
\newblock Medical Physics. 1993;20(2):299-309.

\bibitem{mozumder2020evaluation}
Mozumder M, Tarvainen T.
\newblock Evaluation of temporal moments and {F}ourier transformed data in
  time-domain diffuse optical tomography.
\newblock Journal of the Optical Society of America A. 2020;37(12):1845-56.

\bibitem{schweiger2014toast++}
Schweiger M, Arridge SR.
\newblock The {T}oast++ software suite for forward and inverse modeling in
  optical tomography.
\newblock Journal of Biomedical Optics. 2014;19(4):040801.

\bibitem{Kaipio}
Kaipio JP, Somersalo E.
\newblock Statistical and Computational Inverse Problems.
\newblock Springer, Newyork; 2005.

\bibitem{tarvainen2010}
Tarvainen T, Kolehmainen V, Pulkkinen A, Vauhkonen M, Schweiger M, Arridge SR,
  et~al.
\newblock Approximation error approach for compensating for modelling errors
  between the radiative transfer equation and the diffusion approximation in
  diffuse optical tomography.
\newblock Inverse Problems. 2010;26:015005 (18pp).

\bibitem{Schweiger2005}
Schweiger M, Arridge SR, Nissil{\"a} I.
\newblock Gauss--{N}ewton method for image reconstruction in diffuse optical
  tomography.
\newblock Physics in Medicine \& Biology. 2005;50(10):2365.

\bibitem{tarvainen2008}
Tarvainen T, Vauhkonen M, Arridge SR.
\newblock Gauss-{N}ewton reconstruction method for optical tomography using the
  finite element solution of the radiative transfer equation.
\newblock J Quant Spectrosc Radiat Transf. 2008;109(17-18):2767-78.

\bibitem{rasmussen2006}
Rasmussen CE, Williams CKI.
\newblock Gaussian Processes for Machine Learning.
\newblock the MIT Press, Cambridge, MA; 2006.

\bibitem{pulkkinen2014}
Pulkkinen A, Cox BT, Arridge SR, Kaipio JP, Tarvainen T.
\newblock A {B}ayesian approach to spectral quantitative photoacoustic
  tomography.
\newblock Inverse Problems. 2014;30:065012.

\bibitem{DiNinni2012}
Ninni PD, Bérubé-Lauzière Y, Mercatelli L, Sani E, Martelli F.
\newblock Fat emulsions as diffusive reference standards for tissue simulating
  phantoms?
\newblock Applied Optics. 2012;51(30):7176-82.

\bibitem{Grabtchak2012}
Grabtchak S, Palmer TJ, Foschum F, Liemert A, Kienle A, Whelan WM.
\newblock Experimental spectro-angular mapping of light distribution in turbid
  media.
\newblock Journal of Biomedical Optics. 2012;17(6):067007.

\bibitem{DiNinni2010}
Ninni PD, Martelli F, Zaccanti G.
\newblock The use of India ink in tissue-simulating phantoms.
\newblock Optics Express. 2010;18(26):26854-65.

\bibitem{jacques2013optical}
Jacques SL.
\newblock Optical properties of biological tissues: a review.
\newblock Phys Med Biol. 2013;58(11):R37.

\end{thebibliography}

\end{document}